\documentclass[letterpaper]{raa}           
\usepackage{graphicx,times}
\usepackage{natbib}
\usepackage{amssymb,amsmath}
\usepackage{soul}
\bibpunct{(}{)}{;}{a}{}{,}

\usepackage[pagebackref=true]{hyperref}

\begin{document}

\title{A cosmological full-shape power spectra analysis using pre- and post-reconstructed density fields}

\volnopage{ {\bf 2025} Vol.\ {\bf X} No. {\bf XX}, 000--000}
\setcounter{page}{1}

\author{
Weibing Zhang\inst{1,2}
\and 
Ruiyang Zhao\inst{1,2,3}
\and 
Xiaoyong Mu\inst{1,2}
\and 
Kazuya Koyama\inst{3}
\and 
Ryuichi Takahashi\inst{4}
\and 
Yuting Wang\inst{1,5}
\and 
Gong-Bo Zhao\inst{1,2,5}
}

\institute{
National Astronomical Observatories, Chinese Academy of Sciences, Beijing, 100101, P.R.China; {\it gbzhao@nao.cas.cn}\\
\and
School of Astronomy and Space Sciences, University of Chinese Academy of Sciences, Beijing, 100049, P.R.China\\
\and
Institute of Cosmology and Gravitation, University of Portsmouth, Dennis Sciama Building, Portsmouth PO1 3FX, United Kingdom\\
\and
Faculty of Science and Technology, Hirosaki University, 3 Bunkyo-cho, Hirosaki, Aomori 036-8561, Japan\\
\and
Institute for Frontiers in Astronomy and Astrophysics, Beijing Normal University, Beijing, 102206, P.R.China\\
\vs\no
{\small Received 2025 month day; accepted 2025 month day}
}

\abstract{
In this work, we investigate a joint fitting approach based on theoretical models of power spectra associated with density-field reconstruction. Specifically, we consider the matter auto-power spectra before and after baryon acoustic oscillation (BAO) reconstruction, as well as the cross-power spectrum between the pre- and post-reconstructed density fields. We present redshift-space models for these three power spectra at the one-loop level within the framework of standard perturbation theory (SPT),  and perform a joint analysis using three types of power spectra, and quantify their impact on parameter constraints. When restricting the analysis to wavenumbers \(k \leq 0.2\,h\,\mathrm{Mpc}^{-1}\) and adopting a smoothing scale of \(R_{\mathrm{s}} = 15\,h^{-1}\,\mathrm{Mpc}\), we find that incorporating all three power spectra improves parameter constraints by approximately \(11\%\text{--}16\%\) compared to using only the post-reconstruction power spectrum, with the Figure of Merit (FoM) increasing by \(10.5\%\). These results highlight the advantages of leveraging multiple power spectra in BAO reconstruction, ultimately enabling more precise cosmological parameter estimation.
}

\authorrunning{W. Zhang et al. }           
\titlerunning{A joint power spectra analysis }  
\maketitle

\section{Introduction} \label{sec:intro}

Large galaxy surveys such as the Dark Energy Spectroscopic Instrument (DESI) \citep{DESI:2016fyo} provide vast datasets that are crucial for exploring cosmic large-scale structure. By extracting key cosmological probes, including baryon acoustic oscillations (BAO) \citep{Eisenstein1998, Eisenstein2005, Cole2005} and redshift-space distortions (RSD) \citep{Kaiser:1987qv}, these surveys enable tighter constraints on cosmological parameters, offer insights into the nature of dark energy, and provide a powerful framework for testing alternative theories of gravity.

The coupled photon--baryon fluid leaves a signature on the matter distribution after recombination, appearing as a localized peak in the correlation function or an oscillatory pattern in the power spectrum \citep{Eisenstein:1997gf,Meiksin:1998ra}. The characteristic scale of BAO serves as a standard ruler for cosmological distance measurements \citep{Eisenstein:1998tu}. However, nonlinear structure formation driven by gravity broadens and shifts the BAO peak in the correlation function and damps the oscillations in the power spectrum, leading to a loss of phase coherence and the blurring of BAO measurements \citep{Eisenstein:2006nj,Smith:2007gi,Crocce:2007dt,Seo:2008yx}. Nonlinear evolution of BAO is primarily driven by large-scale bulk flows and gravitational clustering, effects that can be partially corrected via standard reconstruction techniques \citep{Eisenstein:2006nk}. These approaches estimate a displacement field based on the Zel’dovich approximation \citep{zel1970gravitational} and use it to reposition both data and random particles. By separating long-wavelength displacements from the total displacement field, reconstruction effectively transfers crucial information to the reconstructed density field. As a result, standard reconstruction mitigates BAO damping and mode coupling caused by nonlinear evolution, thereby improving measurement precision and reducing systematic shifts \citep{Seo:2008yx,Padmanabhan:2008dd,Seo:2009fp}.

Density-field reconstruction in BAO analysis has motivated deeper investigations into its underlying mechanisms for information recovery. During the process of restoring linear modes contaminated by nonlinear effects, reconstruction transfers higher-order statistical information from the unreconstructed density field, \(\delta^{\mathrm{pre}}\), to the reconstructed density field, \(\delta^{\mathrm{post}}\) \citep{Schmittfull_2015}. In the absence of primordial non-Gaussianity, the higher-order \(N\)-point statistical information in the pre-reconstruction density field arises from gravitationally driven nonlinear evolution. Since reconstruction acts as an approximate inverse process to this evolution, it reduces the non-Gaussianity of the density field, resulting in a more linear and Gaussian post-reconstruction density field \citep{hikage2017perturbation,Hikage:2020fte}. Given these properties, density-field reconstruction can be extended beyond BAO analysis to a wide range of topics, including redshift-space distortions (RSD) \citep{Zhu:2017vtj,Hikage:2019ihj}, neutrino properties \citep{Wang:2022nlx,Zang:2023rpx}, and primordial non-Gaussianity \citep{Shirasaki:2020vkk,Floss:2023ylq,Chen:2024exy}.

Although the post-reconstruction power spectrum, \(P^{\mathrm{post}}\), retains some of the higher-order information from the original unreconstructed density field beyond what is accessible in the pre-reconstruction power spectrum, \(P^{\mathrm{pre}}\), incorporating the cross-power spectrum between the pre- and post-reconstruction fields, \(P^{\mathrm{cross}}\), allows for a more comprehensive extraction of cosmological information. Jointly analyzing the three power spectra, referred to as
\[
P^{\mathrm{all}} = \{ P^{\mathrm{pre}}, P^{\mathrm{post}}, P^{\mathrm{cross}} \},
\]
effectively captures higher-order statistical information \citep{Wang:2022nlx}. Density-field reconstruction enables the transformation of higher-order statistics into two-point statistics, allowing \(P^{\mathrm{post}}\) and \(P^{\mathrm{cross}}\) to be interpreted in terms of specific higher-order statistics of \(\delta^{\mathrm{pre}}\) \citep{Schmittfull_2015,Wang:2022nlx,Sugiyama:2024ggt}. Since these three power spectra reflect different levels of nonlinearity, they exhibit distinct dependencies on cosmological parameters and the higher-order statistics of \(\delta^{\mathrm{pre}}\). A joint analysis of \(P^{\mathrm{all}}\) therefore helps breaking degeneracies among cosmological parameters and small-scale clustering, substantially improving parameter constraints.

Recently, an emulator-based likelihood analysis using galaxy mocks has further demonstrated the effectiveness of this approach, paving the way for its application to observational survey catalogs \citep{prepostEmu}. Besides emulator-based modeling, perturbation theory (PT) can also be employed for the joint analysis of \(P^{\mathrm{all}}\), although modeling smaller scales can be challenging. PT offers valuable insight into the physical underpinnings of this method. Numerous works have developed PT models for the pre-reconstruction power spectrum \citep{Bernardeau:2001qr}, with the effective field theory (EFT) of large-scale structure \citep{Baumann:2010tm,ivanov2022EFT} being widely applied in data analyses \citep[e.g.,][]{Ivanov:2019pdj,DAmico:2019fhj,Zhao:2023ebp,DESI:2024hhd}. Furthermore, various studies have proposed PT-based models for \(P^{\mathrm{post}}\) and \(P^{\mathrm{cross}}\) \citep[e.g.,][]{Padmanabhan:2008dd,Noh_2009,White:2015eaa,Seo:2015eyw,hikage2017perturbation,Chen:2019lpf,Sugiyama:2024qsw,pkcross:2025}.

In this paper, we extend the perturbation-theory-based power spectrum framework of \citet{Hikage:2019ihj} by introducing BAO parameters to account for the Alcock--Paczynski (AP) effect \citep{Alcock:1979mp}. We validate our theoretical models for \(P^{\mathrm{pre}}\), \(P^{\mathrm{post}}\), and \(P^{\mathrm{cross}}\) using \(N\)-body simulation data at redshift \(z = 1.02\). The joint analysis of \(P^{\mathrm{all}}\) yields results consistent with those reported by \citet{prepostEmu}.

This paper is organized as follows. Section~\ref{sec:model} reviews the theoretical model for the power spectra, Section~\ref{sec:validation} presents the joint analysis using $P^{\rm all}$ with simulation data, and Section~\ref{sec:conclusion} summarizes and discusses our main results.

\section{The modeling} \label{sec:model}

In this section, we present one-loop models for the pre-, post-reconstruction, and cross-power spectra of the matter density field in redshift space. Our approach is built upon the effective field theory (EFT) of large-scale structure and includes a leading-order counterterm to account for small-scale (UV) physics. We also incorporate parameters associated with the Alcock--Paczynski (AP) effect to properly model geometric distortions.

To describe the power spectrum as an observable, we begin by defining the matter density contrast:
\begin{equation}
    \delta(\mathbf{x}) \;\equiv\; \frac{\rho(\mathbf{x})}{\bar{\rho}} \;-\; 1,
\end{equation}
where \(\mathbf{x}\) is the comoving coordinate, \(\rho(\mathbf{x})\) is the local matter density, and \(\bar{\rho}\) is the mean density. Under the Newtonian approximation to general relativity, treating matter as a pressureless fluid, the density contrast \(\delta\) and velocity field \(\mathbf{v}\) evolve according to the continuity, Euler, and Poisson equations. Assuming an irrotational velocity field, we introduce the velocity divergence field
{
\begin{equation}
\theta \;=\; -\frac{\nabla \cdot \mathbf{v}}{aH}.
\end{equation}
}
This set of equations can be solved approximately using standard perturbation theory (SPT). In Fourier space, the \(n\)-th order expansions of \(\delta\) and \(\theta\) take the form {\citep[e.g.,][]{Fry:1983cj,Goroff:1986ep,Jain:1993jh,Scoccimarro:1996se,Bernardeau:2001qr,Matsubara_2008}}:
\begin{equation}\label{eq:dn}
    \tilde{\delta}^{(n)}(\mathbf{k})
    \;=\; 
    D^{n}(z)\,
    \int_{\mathbf{k}=\mathbf{q}_{1...n}}
    F_{n}\bigl(\mathbf{q}_{1},\dots,\mathbf{q}_{n}\bigr)\,
    \tilde{\delta}_{\mathrm{L}}(\mathbf{q}_{1})
    \cdots
    \tilde{\delta}_{\mathrm{L}}(\mathbf{q}_{n}),
\end{equation}
{\begin{equation}\label{eq:tn}
    \tilde{\theta}^{(n)}(\mathbf{k})
    \;=\; 
    fD^{n}(z)\,
    \int_{\mathbf{k}=\mathbf{q}_{1...n}}
    G_{n}\bigl(\mathbf{q}_{1},\dots,\mathbf{q}_{n}\bigr)\,
    \tilde{\delta}_{\mathrm{L}}(\mathbf{q}_{1})
    \cdots
    \tilde{\delta}_{\mathrm{L}}(\mathbf{q}_{n}),
\end{equation}
where \(\mathbf{q}_{1...n} = \mathbf{q}_{1} + \cdots + \mathbf{q}_{n}\); \(D(z)\) is the linear growth factor normalized to \(D(z=0)=1\); \(f \equiv \mathrm{d}\ln D/\mathrm{d}\ln a\) denotes the linear growth rate; \(\tilde{\delta}_{\mathrm{L}}\) is the linear density field at \(z=0\);} and \(F_{n}\), \(G_{n}\) are the \(n\)-th order perturbation kernels for the matter density and velocity divergence fields, respectively. We adopt the Einstein--de Sitter approximation \citep{Bernardeau:2001qr}, valid in near-\(\Lambda\)CDM cosmologies, so that {\(D^{(n)}(z)\sim D^{n}(z)\)}. For brevity, we write
\begin{equation}
    \int_{\mathbf{k}=\mathbf{q}_{1...n}}
    \;\equiv\;
    \int 
    \frac{\mathrm{d}\mathbf{q}_{1}\cdots \mathrm{d}\mathbf{q}_{n}}{(2\pi)^{3n-3}}
    \,\delta_{\mathrm{D}}
    \Bigl(\sum_{j=1}^{n}\mathbf{q}_{j}-\mathbf{k}\Bigr),
\end{equation}
where \(\delta_{\mathrm{D}}\) is the Dirac delta function.

To account for redshift-space distortions (RSD) under the distant-observer approximation, we relate real- and redshift-space positions according to the conservation condition. The resulting redshift-space density field is \citep{Matsubara_2008}:
\begin{equation}\label{eq:dz}
    \tilde{\delta}^{\mathrm{z}}(\mathbf{k})
    \;=\;
    \int\!\mathrm{d}\mathbf{x}\,
    e^{-\,i\,\mathbf{k}\cdot\mathbf{x}}\,
    \bigl[\,1+\delta(\mathbf{x})\bigr]\,
    e^{-\,i\,k_{z}\,\frac{v_{z}(\mathbf{x})}{aH}},
    \quad
    \mathbf{k}\neq \mathbf{0},
\end{equation}
where \(k_{z}=\mathbf{k}\cdot \hat{\mathbf{z}}=k\,\mu\) and \(\hat{\mathbf{z}}\) is the unit vector in the line-of-sight direction. Introducing the velocity divergence \(\tilde{\theta}(\mathbf{k})\) and expanding the exponential factor \( e^{-ik_{z} \frac{v_{z}(\mathbf{x})}{aH}} \) in a Taylor series leads to
\begin{equation}\label{eq:evz}
    e^{-ik_{z} \frac{v_{z}(\mathbf{x})}{aH}} = \sum_{n=0}^{\infty} \frac{(k\mu)^{n}}{n!} \int \frac{\mathrm{d}\mathbf{q}_{1} \cdots \mathrm{d}\mathbf{q}_{n}}{(2\pi)^{3n}} \prod_{m=1}^{n} \frac{\mu_{m}}{q_{m}} \tilde{\theta}(\mathbf{q}_{m}) e^{i\mathbf{q}_{m} \cdot \mathbf{x}},
\end{equation}
where \(\mu_m=\mathbf{q}_m\cdot\hat{\mathbf{z}}/q_m\). 
By substituting Eq.~(\ref{eq:evz}) into Eq.~(\ref{eq:dz}) and performing a perturbative expansion of the density field $\delta$ and velocity divergence field $\theta$, one can derive the \(n\)-th order density fluctuation in redshift space as:
\begin{equation}\label{eq:dnz}
    \tilde{\delta}^{(n)}(\mathbf{k})
    \;=\;
    D^{n}(z)\,
    \int_{\mathbf{k}=\mathbf{q}_{1...n}}
    Z_{n}\bigl(\mathbf{q}_{1},\dots,\mathbf{q}_{n}\bigr)\,
    \tilde{\delta}_{\mathrm{L}}(\mathbf{q}_{1})
    \cdots
    \tilde{\delta}_{\mathrm{L}}(\mathbf{q}_{n}),
\end{equation}
where \(Z_{n}\) is the \(n\)-th order redshift-space kernel \citep[e.g.,][]{Heavens:1998es,Scoccimarro:1999ed,matsubara2008nonlinear,Hikage:2019ihj}. In what follows, we omit the superscript ``\(\mathrm{z}\)''.

We apply the standard reconstruction technique \citep{Eisenstein:2006nk}, which estimates a shift field \(\tilde{\mathbf{s}}\) from the smoothed nonlinear density field \(\tilde{\delta}\) using the negative Zel’dovich approximation \citep{zel1970gravitational}:
\begin{equation}
    \tilde{\mathbf{s}}(\mathbf{k})
    \;=\;
    -\,i\,\frac{\mathbf{k}}{k^{2}}\,W(k)\,\tilde{\delta}(\mathbf{k}),
    \quad
    W(k)\;=\;\exp(-\,k^{2}\,R_{\mathrm{s}}^{2}/2),
\end{equation}
{with \(R_{\mathrm{s}}\) denoting the smoothing scale used in reconstruction.} This procedure mitigates nonlinear effects from bulk flows and cluster formation. Displacing both data and random particles by \(\tilde{\mathbf{s}}(\mathbf{k})\) yields the displaced and shifted fields \(\tilde{\delta}^{(\mathrm{d})}\) and \(\tilde{\delta}^{(\mathrm{s})}\), whose difference defines the reconstructed density field:
\begin{equation}
    \tilde{\delta}^{(\mathrm{rec})}
    \;\equiv\;
    \tilde{\delta}^{(\mathrm{d})}
    \;-\;
    \tilde{\delta}^{(\mathrm{s})}.
\end{equation}
Because the matter sample has a sufficiently high number density, discreteness effects in reconstruction \citep{Sugiyama:2024ggt} are negligible.

An analogous perturbative expansion exists for \(\tilde{\delta}^{(\mathrm{rec})}\):
\begin{equation}
    \tilde{\delta}^{(\mathrm{rec},n)}(\mathbf{k})
    \;=\;
    D^{n}(z)\,
    \int_{\mathbf{k}=\mathbf{q}_{1...n}}
    Z_{n}^{(\mathrm{rec})}\bigl(\mathbf{q}_{1},\dots,\mathbf{q}_{n}\bigr)\,
    \tilde{\delta}_{\mathrm{L}}(\mathbf{q}_{1})
    \cdots
    \tilde{\delta}_{\mathrm{L}}(\mathbf{q}_{n}),
\end{equation}
where \(Z_{n}^{(\mathrm{rec})}\) is given by \citet{Hikage:2019ihj,Hikage:2020fte}.

Having obtained perturbative forms for the density fields before and after reconstruction, we can compute the redshift-space power spectrum
\begin{equation}
    \langle 
    \tilde{\delta}(\mathbf{k})\,\tilde{\delta}(\mathbf{k}')
    \rangle
    \;=\;
    (2\pi)^3\,
    \delta_{\mathrm{D}}(\mathbf{k}+\mathbf{k}')\,
    P(\mathbf{k}).
\end{equation}
Up to one-loop order, the pre-reconstruction power spectrum is
\begin{equation}\label{eq:ph}
    P_{1\text{-loop}}(k,\mu)
    \;=\;
    D^{2}(z)\,P_{11}(k,\mu)
    \;+\;
    D^{4}(z)\,\bigl[
        P_{22}(k,\mu)
        \;+\;
        P_{13}(k,\mu)
    \bigr],
\end{equation}
where
\begin{equation}
    P_{11}(k,\mu)
    \;=\;
    (1 + f\,\mu^2)^{2}\,P_{\mathrm{L}}(k),
\end{equation}
\begin{equation}
    P_{22}(k,\mu)
    \;=\;
    2\,
    \int
    \frac{\mathrm{d}\mathbf{q}}{(2\pi)^{3}}\,
    \bigl[
       Z_{2}(\mathbf{k}-\mathbf{q},\mathbf{q})
    \bigr]^{2}\,
    P_{\mathrm{L}}\bigl(\lvert\mathbf{k}-\mathbf{q}\rvert\bigr)\,
    P_{\mathrm{L}}(q),
\end{equation}
\begin{equation}
    P_{13}(k,\mu)
    \;=\;
    6\,(1 + f\,\mu^2)\,P_{\mathrm{L}}(k)\,
    \int
    \frac{\mathrm{d}\mathbf{q}}{(2\pi)^{3}}\,
    Z_{3}\bigl(\mathbf{k},\mathbf{q},-\mathbf{q}\bigr)\,
    P_{\mathrm{L}}(q),
\end{equation}
{\(P_{\mathrm{L}}(k)\) is the linear power spectrum evaluated at \(z=0\).} Similar expressions hold for the one-loop post-reconstruction power spectrum and the cross-power spectrum, but the second-order and third-order kernels \((Z_{2}, Z_{3})\) are replaced by their post- and cross-reconstruction counterparts \(\bigl(Z_{2}^{(\mathrm{rec})}, Z_{3}^{(\mathrm{rec})}\bigr)\) and \(\bigl(Z_{2}^{(\mathrm{x})}, Z_{3}^{(\mathrm{x})}\bigr)\) \citep{pkcross:2025}.

Theoretical predictions are often expanded into Legendre multipoles:
\begin{equation}
    P_{\ell}(k)
    \;=\;
    \frac{2\ell+1}{2}
    \int_{-1}^{1}\!\mathrm{d}\mu\,
    P(k,\mu)\,\mathcal{P}_{\ell}(\mu),
\end{equation}
where \(\mathcal{P}_{\ell}(\mu)\) is the Legendre polynomial of order \(\ell\). One-loop SPT alone does not fully capture nonlinear small-scale physics, so EFT introduces a counterterm to absorb UV contributions. For the one-loop pre-reconstruction spectrum, the counterterm is proportional to \(k^2P_{\mathrm{L}}(k)\) \citep{Senatore:2014vja}. We adopt a similar form for the post-reconstruction and cross-power spectra \citep{Hikage:2019ihj,pkcross:2025}:
\begin{equation}
    P_{\mathrm{ct},\ell}(k)
    \;=\;
    c_{\ell}\,k^{2}\,P_{\ell}^{\mathrm{L}}(k).
\end{equation}

Converting redshift to comoving distance introduces the Alcock--Paczynski (AP) effect \citep{Alcock:1979mp}, which distorts \((k,\mu)\) when the fiducial cosmology differs from the true one. Labeling true-cosmology coordinates by \((k',\mu')\), the full power spectrum with counterterms reads
\begin{equation}
    P^{\mathrm{theory}}(k',\mu')
    \;=\;
    P_{1\text{-loop}}(k',\mu') \;+\; P_{\mathrm{ct}}(k',\mu'),
\end{equation}
and we recover the 2D power spectrum from its multipoles for \(\ell=0,2,4\). Higher-order multipoles are negligible for the scales of interest. The AP effect rescales \(k\) and \(\mu\) according to
\[
    k'
    \;=\;
    \frac{k}{\alpha_{\perp}}\,
    \sqrt{1\,+\,\mu^2\bigl(F^{-2}-1\bigr)},
    \quad
    \mu'
    \;=\;
    \frac{\mu/F}{
    \sqrt{1\,+\,\mu^2\bigl(F^{-2}-1\bigr)}
    },
    \quad
    F
    \;=\;
    \frac{\alpha_{\parallel}}{\alpha_{\perp}},
\]
where
\[
    \alpha_{\parallel}
    \;=\;
    \frac{H^{\mathrm{f}}(z_{\mathrm{eff}})\,r_{\mathrm{d}}^{\mathrm{f}}}{
          H(z_{\mathrm{eff}})\,r_{\mathrm{d}}},
    \quad
    \alpha_{\perp}
    \;=\;
    \frac{D_{\mathrm{A}}(z_{\mathrm{eff}})\,r_{\mathrm{d}}^{\mathrm{f}}}{
          D_{\mathrm{A}}^{\mathrm{f}}(z_{\mathrm{eff}})\,r_{\mathrm{d}}},
\]
relate AP parameters to Hubble function \( H \) and the angular diameter
distance \( D_{\mathrm{A}} \) at the effective redshift \( z_{\mathrm{eff}} \), and \(r_{\mathrm{d}}\) is the sound horizon at the drag epoch. Note that the superscript \( \mathrm{f} \) indicates fiducial values. Discrete \(k\)-binning in measurements is accounted for by averaging the theoretical predictions over each bin:
\begin{equation}
    P_{\ell}^{\mathrm{th}}(k_{\mathrm{bin}})
    \;\approx\;
    \frac{\displaystyle
          \int_{k_{\mathrm{bin,min}}}^{k_{\mathrm{bin,max}}}
          \mathrm{d}k\;\!k^{2}\,P_{\ell}^{\mathrm{AP}}(k)
         }{\displaystyle
          \int_{k_{\mathrm{bin,min}}}^{k_{\mathrm{bin,max}}}
          \mathrm{d}k\;\!k^{2}},
\end{equation}
where \(P_{\ell}^{\mathrm{AP}}(k)\) includes the AP transformation. This procedure allows a more accurate comparison between theory and binned data \(P_{\ell}^{\mathrm{data}}(k_{\mathrm{bin}})\) in subsequent likelihood analyses.

\section{Results} \label{sec:validation}

In this section, we present our analysis using a suite of high-resolution \(N\)-body simulations. This mock dataset was previously used in 
\citet{Hikage:2020fte, pkcross:2025}. The input cosmological parameters for these simulations are based on the best-fit values from the Planck 2015 TT, TE, EE+lowP measurements: 
\(\Omega_b = 0.0492\), \(\Omega_m = 0.3156\), \(h = 0.6727\), \(n_s = 0.9645\), and \(\sigma_8 = 0.831\) \citep{Planck:2015fie}. 

The initial linear power spectrum is computed using \texttt{CAMB} \citep{Lewis:1999bs}, which is also used to calculate our theoretical model predictions. The initial mass particle distribution is generated with second-order Lagrangian perturbation theory (2LPT) \citep{Crocce:2006ve, Nishimichi:2008ry}. We then perform \(N\)-body simulations using \texttt{Gadget-2} \citep{Springel:2005mi} to generate 4000 realizations, each with a box size of \(L = 500\,h^{-1}\mathrm{Mpc}\), containing \(512^3\) particles at redshift \(z = 1.02\). After standard reconstruction is applied to each realization, the multipoles of the pre-reconstruction, post-reconstruction, and cross-power spectra are measured from the simulation samples. Owing to the large measurement uncertainties associated with the hexadecapole, only monopole and quadrupole measurements are used in this work. Further details on the simulation data and reconstruction procedure can be found in \citet{Hikage:2020fte}.

We measure the power spectrum multipole components from 4000 realizations of our simulations and use them to estimate the covariance matrix:
\begin{equation}
    \mathbf{Cov}_{\ell,\ell^{\prime}}(k,k^{\prime}) = \frac{1}{N-1}
    \sum_{i=1}^{N} \bigl[P_{\ell,i}(k) - \bar{P}_{\ell}(k)\bigr]\,
    \bigl[P_{\ell^{\prime},i}(k^{\prime}) - \bar{P}_{\ell^{\prime}}(k^{\prime})\bigr],
    \label{eq:cov}
\end{equation}
where \(N = 4000\) is the total number of realizations, and 
\(\bar{P}_{\ell}(k) = \tfrac{1}{N}\sum_{i=1}^{N} P_{\ell,i}(k)\) is the mean power spectrum multipole across all realizations.

Due to the limited box size \(L=500\,h^{-1}\mathrm{Mpc}\), the mean power spectrum \(\bar{P}_{\ell}\) exhibits a pronounced sawtooth pattern on large scales, especially for \(\ell\ge2\). To mitigate this effect, we use an additional set of 8 realizations with a larger volume, \(V=(4\,h^{-1}\mathrm{Gpc})^3\), to correct the \(V=(500\,h^{-1}\mathrm{Mpc})^3\) data following \citet{pkcross:2025}:
\begin{equation}
    \bar{P}_{\ell}^{(\mathrm{T}),4h^{-1}\mathrm{Gpc}} \;=\; 
    \frac{\bar{P}_{\ell}^{(\mathrm{pre}),4h^{-1}\mathrm{Gpc}}}
         {\bar{P}_{\ell}^{(\mathrm{pre}),500h^{-1}\mathrm{Mpc}}}
    \;\bar{P}_{\ell}^{(\mathrm{T}),500h^{-1}\mathrm{Mpc}},
\end{equation}
where \(\mathrm{T}\in\{\mathrm{pre},\,\mathrm{post},\,\mathrm{cross}\}\) and the superscripts \(4h^{-1}\mathrm{Gpc}\) or \(500h^{-1}\mathrm{Mpc}\) indicate the side length of the simulation box. These larger-volume simulations are run with the same cosmological parameters as the \(500\,h^{-1}\mathrm{Mpc}\) boxes \citep{Hikage:2020fte}. This “grid correction” ensures an accurate representation of large-scale modes, thereby improving our subsequent likelihood analysis.

The likelihood function is given by
\begin{equation}
    \mathcal{L} \;=\; \exp\!\Bigl(-\tfrac{\chi^{2}}{2}\Bigr),
\end{equation}
where the chi-square statistic takes the form
\begin{equation}
    \chi^{2}(\mathbf{p}) 
    \;=\; \sum_{\ell,\ell^{\prime}=0,2}
    \sum_{\substack{i,j \\ k_{\min}\le k_{i},k_{j}\le k_{\max}}}
    \Bigl[P_{\ell}^{\mathrm{th}}(k_{i}; \mathbf{p})
          \;-\; P_{\ell}^{\mathrm{sim}}(k_{i})\Bigr]\,
    \mathbf{Cov}_{\ell,\ell^{\prime}}^{-1}(k_{i},k_{j})\,
    \Bigl[P_{\ell^{\prime}}^{\mathrm{th}}(k_{j}; \mathbf{p})
          \;-\; P_{\ell^{\prime}}^{\mathrm{sim}}(k_{j})\Bigr].
\end{equation}
Since the number of realizations \(N\) is finite, the inverse of the covariance matrix is rescaled by the Hartlap factor \((N - N_{\mathrm{bin}} - 2)/(N - 1)\) \citep{Hartlap:2006kj}, where \(N_{\mathrm{bin}}\) is the number of bins used in the fit.

In each power spectrum model, we treat five parameters as free:
\begin{equation}
    \mathbf{p} \;=\; \{\alpha_{\parallel},\,\alpha_{\perp},\,f,\,c_{0},\,c_{2}\}.
\end{equation}
In the joint fit using three types of power spectra, the parameters \(\alpha_{\parallel}\), \(\alpha_{\perp}\), and \(f\) are shared across all three models, while each model retains its own counterterm parameters \((c_{0},c_{2})\). This results in 9 total free parameters. All parameters have uniform (flat) priors, \(\mathcal{U}(\min,\max)\), whose ranges are listed in Table~\ref{prior}.

\begin{table}[ht]
    \centering
    \caption{Priors for free parameters.}
    \label{prior}
    \begin{tabular}{c c c c c c c}
      \hline
      \hline
      Parameter & \( \alpha_{\parallel} \) & \( \alpha_{\perp} \) & \( f \) & \( c_0 \) & \( c_2 \) \\
      \hline
      Prior & \( \mathcal{U}(0.6, 1.4) \) & \( \mathcal{U}(0.6, 1.4) \) & \( \mathcal{U}(0.2, 1.5) \) & \( \mathcal{U}(-10^3, 10^3) h^{-2} \mathrm{Mpc}^2 \) & \( \mathcal{U}(-10^3, 10^3) h^{-2} \mathrm{Mpc}^2 \) \\
      \hline
    \end{tabular}
\end{table}

We construct our parameter estimation pipeline within the \texttt{Cobaya} framework \citep{Torrado:2020dgo}, using Markov Chain Monte Carlo (MCMC) sampling \citep{Lewis:2002ah} to explore the posterior distributions. The resulting MCMC chains are analyzed with \texttt{GetDist} \citep{Lewis:2019xzd}, which yields marginalized posteriors and the corresponding contour plots. To ensure convergence, we require the Gelman--Rubin statistic to satisfy \(R - 1 < 0.001\). In addition, we use \texttt{iminuit} \citep{iminuit,James:1975dr} to minimize \(\chi^2\) and obtain the best-fit parameter values.
 
{In standard reconstruction, the choice of smoothing scale  is essential. A larger \(R_{\mathrm{s}}\) reduces the effectiveness of reconstruction, while a smaller \(R_{\mathrm{s}}\) may enhance the gain compared to the pre-reconstructed case. However, excessively small values (e.g., \(R_{\mathrm{s}} < 10h^{-1}\mathrm{Mpc}\)) can introduce large-scale nonlinearities due to reconstruction inaccuracy\citep{hikage2017perturbation}, making theoretical predictions more challenging and potentially compromising the reliability of the results. To ensure robustness, we adopt \(R_{\mathrm{s}} = 15h^{-1}\mathrm{Mpc}\) in this work.
Our likelihood analysis includes both individual and joint fits of the three power spectra.}
In individual fits, we set \(k_{\min} = 0.02\,h\,\mathrm{Mpc}^{-1}\). For our main results, we adopt \(k_{\max} = 0.20\,h\,\mathrm{Mpc}^{-1}\), while also testing a more conservative choice of \(k_{\max} = 0.16\,h\,\mathrm{Mpc}^{-1}\). The corresponding results for this more conservative limit are presented in Appendix~\ref{sec:appendix}.

\begin{figure}
\centering
\includegraphics[width=0.49\textwidth]{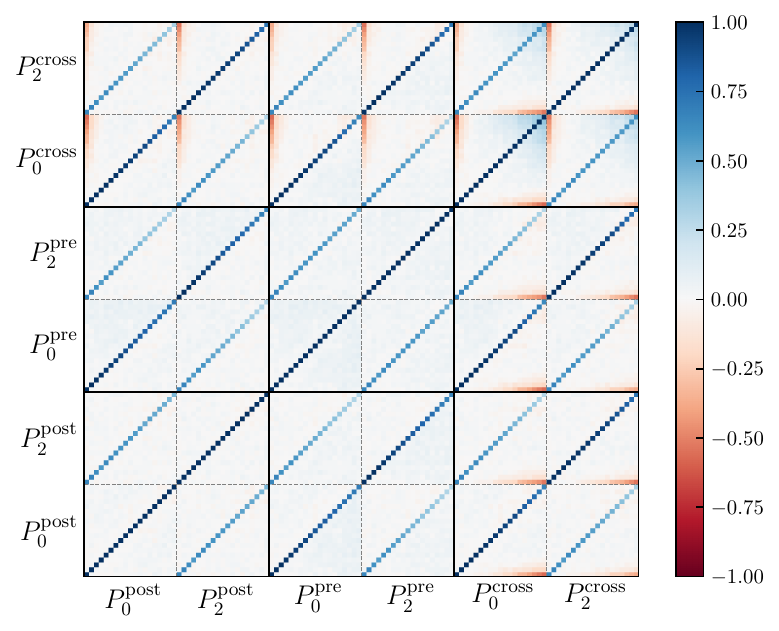}
\includegraphics[width=0.49\textwidth]{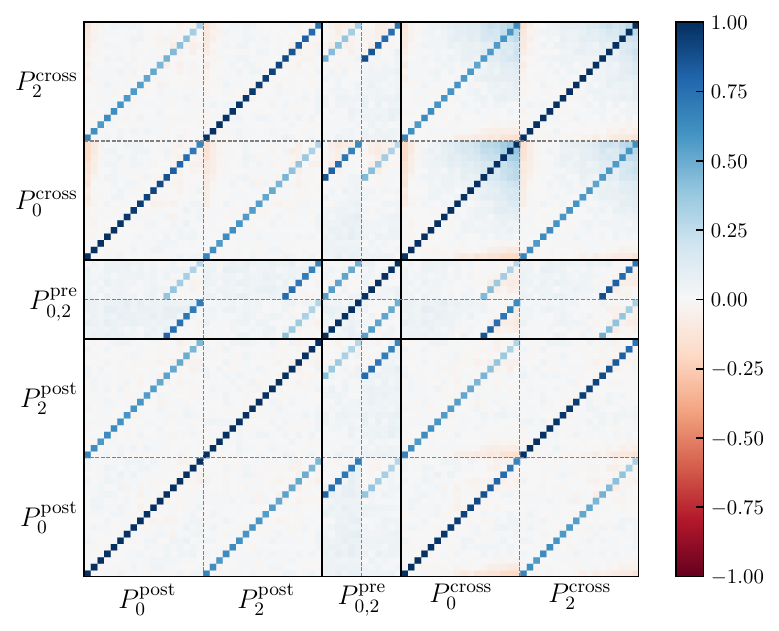}
\caption{\label{fig:corrmat}
Correlation coefficient matrices derived from the covariance matrix in Eq.~(\ref{eq:cov}). 
The reconstruction is performed with a smoothing scale \(R_{\mathrm{s}} = 15h^{-1}\,\mathrm{Mpc}\), 
using modes up to \(k_{\mathrm{max}} = 0.20\,h\,\mathrm{Mpc}^{-1}\). 
\emph{Left panel}: the full matrix, with \(k_{\min}=0.01\,h\,\mathrm{Mpc}^{-1}\) for all three power spectra.
\emph{Right panel}: a modified setup where \(k_{\min}^{\mathrm{pre}}=0.14\,h\,\mathrm{Mpc}^{-1}\) 
and \(k_{\min}^{\mathrm{post}}=k_{\min}^{\mathrm{cross}}=0.02\,h\,\mathrm{Mpc}^{-1}\).}
\end{figure}

\begin{figure}[hbtp]
\centering
\includegraphics[width=0.8\textwidth]{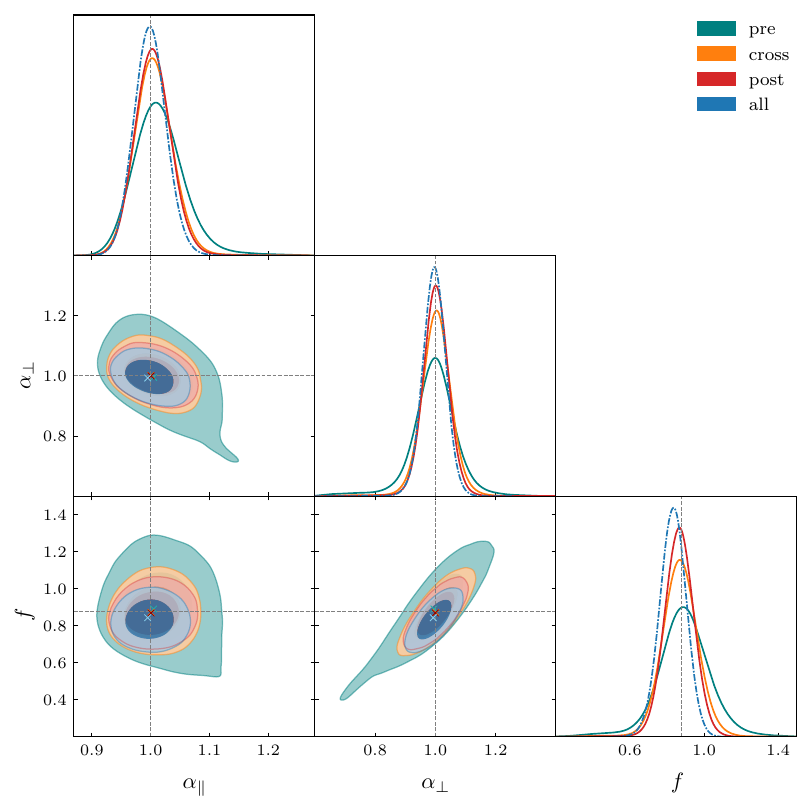}
\caption{\label{fig:contour_0p200}
Constraints on the parameters \(\{\alpha_\parallel, \alpha_\perp, f\}\) obtained by fitting the theoretical model to the monopole and quadrupole of the pre-reconstruction (\emph{teal}), post-reconstruction (\emph{red}), and cross-power (\emph{orange}) spectra, as well as all three simultaneously (\emph{blue}). The crosses mark the best-fit parameters, and the dashed lines show the fiducial values. The reconstruction is performed with \(R_{\mathrm{s}}=15\,h^{-1}\mathrm{Mpc}\). In individual fits, the wavenumber range is \(0.02<k<0.20\,h\,\mathrm{Mpc}^{-1}\). For the joint fit (\(P^{\mathrm{all}}\)), we choose \(k_{\min}^{\mathrm{pre}}=0.14\,h\,\mathrm{Mpc}^{-1}\) and \(k_{\min}^{\mathrm{post,cross}}=0.02\,h\,\mathrm{Mpc}^{-1}\), with \(k_{\max}=0.20\,h\,\mathrm{Mpc}^{-1}\).}
\end{figure}

\begin{table}[htbp]
\centering
\caption{Best-fit values, 68\% confidence intervals, and Figures of Merit (FoM) for the various fit configurations shown in Fig.~\ref{fig:contour_0p200}, assuming \(k_{\max}=0.20\,h\,\mathrm{Mpc}^{-1}\). 
The last column indicates the relative reduction in parameter uncertainties and the increase in FoM for the joint fit \(P^{\mathrm{all}}\) compared to the post-reconstruction fit \(P^{\mathrm{post}}\). The fiducial values of the parameters are \(\alpha_{\parallel}^{\mathrm{f}} = \alpha_{\perp}^{\mathrm{f}} = 1 \) and \(f_{\mathrm{in}} = 0.8796 \).}
\label{table_0p200}
\begin{tabular}{l c c c c c}
\hline\hline
\multicolumn{1}{c}{Case} & 
\(\;P^{\mathrm{pre}}\;\) & 
\(\;P^{\mathrm{post}}\;\) & 
\(\;P^{\mathrm{cross}}\;\) & 
\(\;P^{\mathrm{all}}\;\) & 
\(\;1 - \sigma^\mathrm{all}/\sigma^\mathrm{post}\) \\
\hline
\(\alpha_\parallel\) 
& \(1.016^{+0.035}_{-0.047}\) 
& \(1.005^{+0.028}_{-0.032}\) 
& \(1.006^{+0.029}_{-0.034}\) 
& \(0.999 \pm 0.028\) 
& \(11\%\) \\[6pt]

\(\alpha_\perp\) 
& \(0.995 \pm 0.082\) 
& \(1.001 \pm 0.047\) 
& \(1.003 \pm 0.056\) 
& \(0.996 \pm 0.040\) 
& \(16\%\) \\[6pt]

\(f\) 
& \(0.89 \pm 0.15\) 
& \(0.865 \pm 0.084\) 
& \(0.87 \pm 0.10\) 
& \(0.832 \pm 0.073\) 
& \(14\%\) \\[6pt]

\hline
\text{FoM} 
& \(17.3\) 
& \(24.6\) 
& \(22.4\) 
& \(27.2\) 
& \(10.5\%\) \\
\hline
\end{tabular}
\end{table}

In the joint analysis of \(P^{\mathrm{all}} = \{P^{\mathrm{pre}},\,P^{\mathrm{post}},\,P^{\mathrm{cross}}\}\), we initially attempted a combined fit using all data points with \(k \in (0.02,\,0.20)\,h\,\mathrm{Mpc}^{-1}\). However, because all three power spectra closely resemble the linear power spectrum on large scales, their correlation coefficients (the correlation among $P^{\mathrm{pre}},\,P^{\mathrm{post}}$ and $P^{\mathrm{cross}}$ at the same $k$) approaches unity, as seen in the left panel of Fig.~\ref{fig:corrmat}. This means that including all three power spectra in the analysis is redundant, which can cause numerical instabilities when we invert the data covariance matrix.

To address this issue, we remove part of the large-scale modes from the pre-reconstruction power spectrum, which effectively reduces the redundancy in the data vector. In practice, we set 
\(k_{\min}^{\mathrm{pre}} = 0.14\,h\,\mathrm{Mpc}^{-1}\) 
and 
\(k_{\min}^{\mathrm{post}} = k_{\min}^{\mathrm{cross}} = 0.02\,h\,\mathrm{Mpc}^{-1}\). 
The right panel of Fig.~\ref{fig:corrmat} shows the correlation coefficients after this cut, whose absolute values are all well below unity, ensuring a stable numeric inversion in the likelihood analysis.

Figure~\ref{fig:contour_0p200} presents the resulting constraints on BAO and RSD parameters obtained from each individual power spectrum \(\bigl(P^{\mathrm{pre}}, P^{\mathrm{post}}, P^{\mathrm{cross}}\bigr)\) and their combined fit \(P^{\mathrm{all}}\). The best-fit results are all consistent with the fiducial parameter values within the 68\% confidence region. Among the individual fits, \(P^{\mathrm{post}}\) delivers the tightest constraints, with \(P^{\mathrm{cross}}\) performing comparably well. A complete summary of these results is provided in Table~\ref{table_0p200}.

Table~\ref{table_0p200} also demonstrates how the joint fit \(P^{\mathrm{all}} = \{P^{\mathrm{pre}}, P^{\mathrm{post}}, P^{\mathrm{cross}}\}\) substantially tightens the parameter constraints relative to the post-reconstruction power spectrum alone, offering reductions in \(\sigma_p\) of up to \(14\%\). The corresponding Figure of Merit (FoM) for the BAO and RSD parameters, defined as
\begin{equation}
    \text{FoM} \;=\; \left[\frac{1}{\det(\mathbf{C})}\right]^{\!\!1/(2N_p)},
\end{equation}
also shows a \(10.5\%\) enhancement when combining all three power spectra, where \(\mathbf{C}\) is the covariance matrix of \(\{\alpha_{\parallel},\,\alpha_{\perp},\,f\}\) estimated from the Markov Chain Monte Carlo (MCMC) samples, and \(N_p=3\) denotes the number of free parameters in \(\mathbf{C}\). This underscores the value of including all available spectra to achieve tighter cosmological parameter constraints.

\begin{figure}[htbp]
\centering
\includegraphics[width=0.48\textwidth]{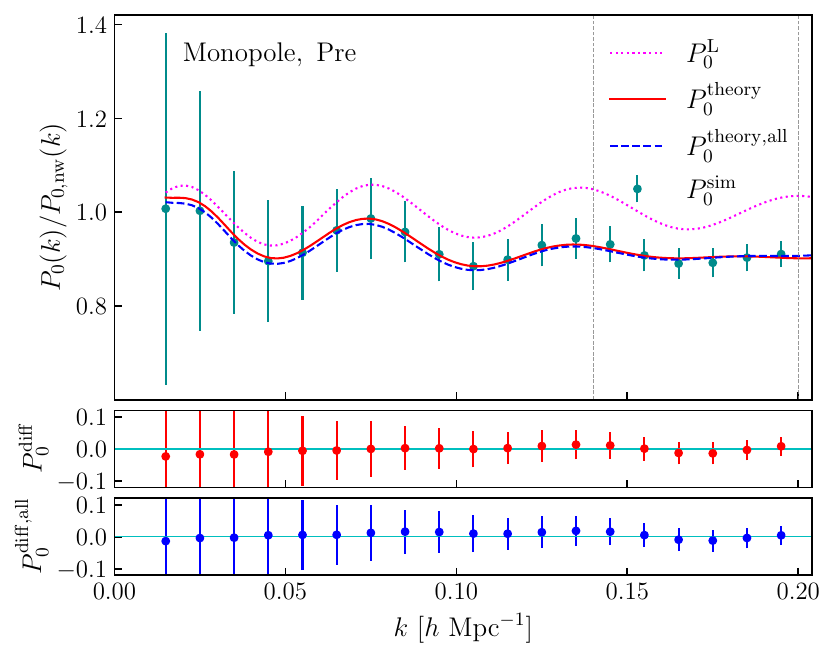}
\includegraphics[width=0.48\textwidth]{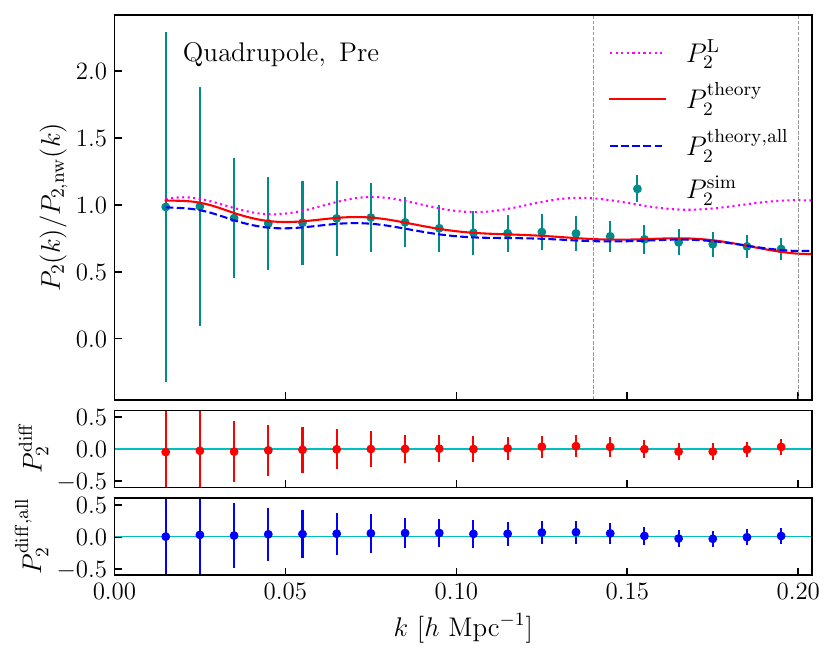}

\includegraphics[width=0.48\textwidth]{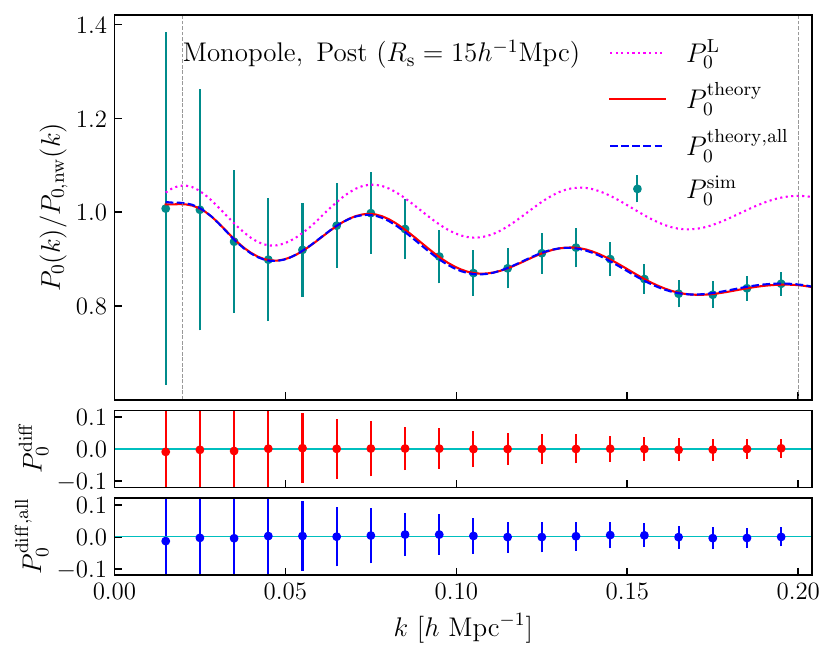}
\includegraphics[width=0.48\textwidth]{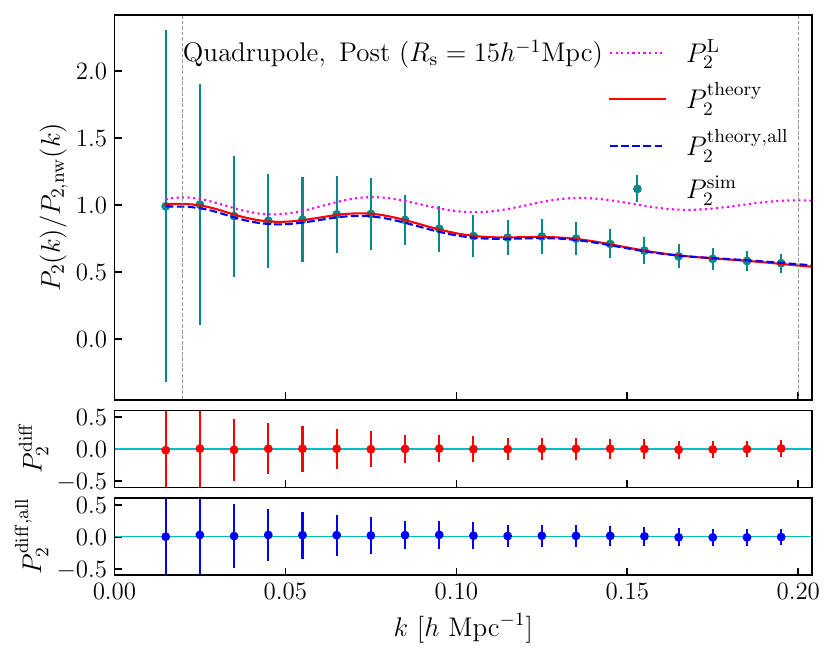}

\includegraphics[width=0.48\textwidth]{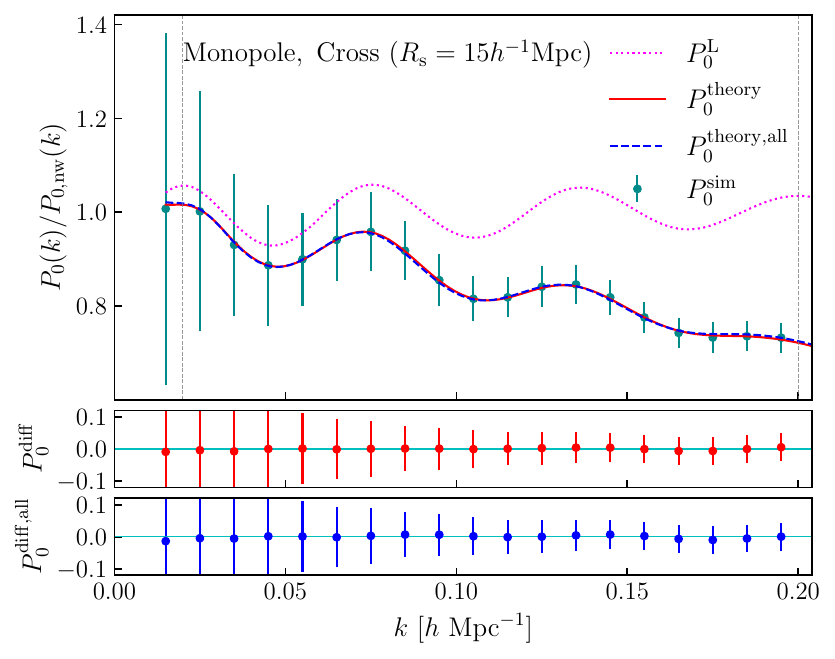}
\includegraphics[width=0.48\textwidth]{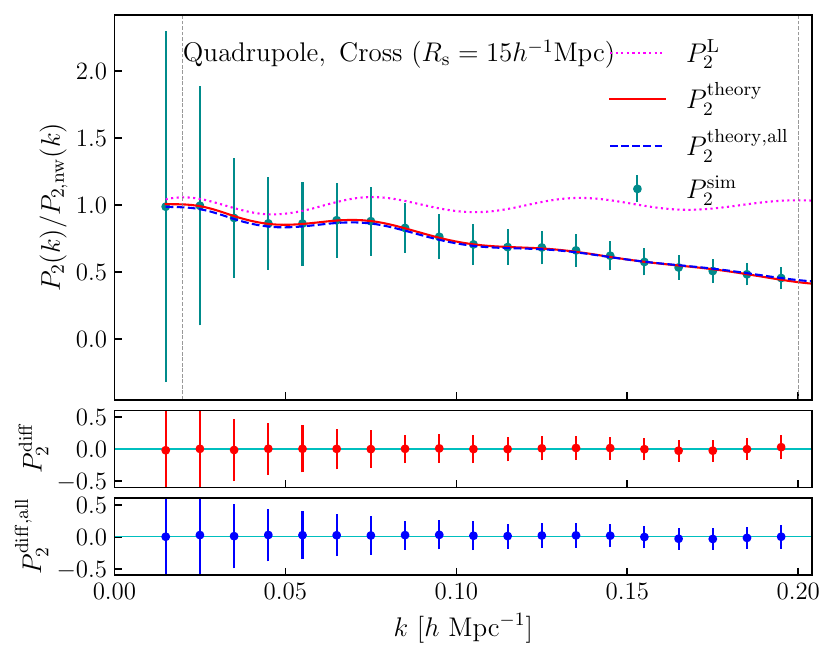}

\caption{\label{fig:wiggle_plk_0p200}
Comparison of the measured and modeled power spectrum monopole (\(\ell=0\), left column) and quadrupole (\(\ell=2\), right column) for pre-reconstruction (top panels), post-reconstruction (middle panels), and cross-power (bottom panels). The teal points with error bars denote the simulation results and their standard deviations. The red solid curves are the best-fit theoretical predictions from individual fits of each power spectrum, whereas the blue dashed curves display the predictions from the joint fit of all three spectra. The magenta dotted lines show the linear theory power spectra computed from the fiducial cosmology. 
All power spectra are divided by the corresponding no-wiggle linear reference. 
The bottom portion of each subplot shows 
\(P_{\ell}^{\mathrm{diff}} \equiv P_{\ell}^{\mathrm{sim}}/P_{\ell}^{\mathrm{theory}} - 1\), 
the relative difference between the simulations and the theoretical prediction. 
The vertical gray dashed lines mark the lower and upper \(k\)-cuts used in the joint fits.}
\end{figure}

Using the best-fit parameters obtained from each fit, we compute the corresponding theoretical monopole and quadrupole power spectra and compare them with the simulation data in Fig.~\ref{fig:wiggle_plk_0p200}. We decompose the linear power spectrum into wiggle and no-wiggle components via a polynomial-based method \citep{Hinton:2016atz}, writing
\(\,P_{\mathrm{L}}(k) = P_{\mathrm{w}}(k) + P_{\mathrm{nw}}(k)\).
Applying the fiducial linear growth rate \(f_{\mathrm{in}} = 0.8796\), we calculate the linear redshift-space power spectrum and its multipoles using the Kaiser formula \citep{Kaiser:1987qv}:
\begin{equation}
    P^{\mathrm{L}}(k,\mu) \;=\; \bigl(1 + f_{\mathrm{in}}\,\mu^{2}\bigr)^{2}\,P_{\mathrm{L}}(k).
\end{equation}
For ease of comparison, both theoretical and simulation power spectra are divided by \(P_{\ell,\mathrm{nw}}\). 
Figure~\ref{fig:wiggle_plk_0p200} also shows the corresponding linear monopole and quadrupole; 
as expected, BAO wiggles become more prominent in the post-reconstruction and cross-power spectra 
than in the pre-reconstruction spectrum. Meanwhile, the cross-power spectrum exhibits a gradually 
decreasing amplitude, owing to the absence of IR cancellation \citep{Sugiyama:2024eye}.

As seen in Fig.~\ref{fig:wiggle_plk_0p200}, the monopole and quadrupole from each individually fitted model 
agree closely with the simulation data, and the joint fit (\(P^\mathrm{all}\)) also provides consistent results. 
Notably, for both post-reconstruction and cross-power spectra, the joint-fit curves are similar to those obtained in the individual fits. However, the BAO wiggles in \(P^\mathrm{pre}\) are not perfectly captured by the current perturbation theory approach. Incorporating IR resummation \citep{Sugiyama:2024eye, Sugiyama:2024qsw} could further improve the BAO modeling, and we plan to pursue this extension in future work.

\section{Conclusion and discussions} \label{sec:conclusion} 

In this paper, we conduct a full-shape analysis for BAO and RSD parameters using the power spectra derived from both pre- and post-reconstruction density fields, while also including their cross-power spectrum. Our theoretical framework is developed at one-loop order in redshift space for the matter density field. We incorporate the Alcock--Paczyński (AP) effect into our models to constrain BAO parameters.

To extend standard perturbation theory (SPT) to smaller scales, we introduce effective field theory (EFT) counterterms to account for unmodeled ultraviolet (UV) physics. However, the counterterms may partially degenerate with infrared (IR) contributions and other modeling uncertainties. Our theoretical models for \(P^{\mathrm{pre}}\), \(P^{\mathrm{post}}\), and \(P^{\mathrm{cross}}\) thus fit simulation data for the power spectrum multipoles, constraining the BAO parameters \(\alpha_{\parallel}, \alpha_{\perp}\) and the linear growth rate \(f\). Each of these three models provides unbiased parameter estimates individually up to \(k_{\max}=0.2\,h\,\mathrm{Mpc}^{-1}\), assuming a reconstruction smoothing scale \(R_{\mathrm{s}} = 15\,h^{-1}\mathrm{Mpc}\) and a covariance matrix estimated from 4000 \(N\)-body realizations at redshift \(z=1.02\), each with a volume of \(\bigl(500\,h^{-1}\mathrm{Mpc}\bigr)^3\). Among these models, \(P^{\mathrm{post}}\) yields the tightest constraints, while \(P^{\mathrm{cross}}\) is intermediate between \(P^{\mathrm{pre}}\) and \(P^{\mathrm{post}}\), yet quite close to the latter in terms of constraining power.

We carry out a joint fit of all three power spectra, \(P^{\mathrm{all}} = \{P^{\mathrm{pre}},\,P^{\mathrm{post}},\,P^{\mathrm{cross}}\}\), to further tighten parameter constraints. Because these spectra are strongly correlated on large scales, especially \(P^{\mathrm{pre}}\) and \(P^{\mathrm{post}}\), numerical instability in the covariance matrix can arise when using limited-volume simulations and a finite number of realizations. In addition, small-scale nonlinearities may also bias the inferred parameters if not fully captured by the perturbation theory. We employ a pragmatic remedy in this work: increasing the minimum wavenumber \(k_{\min}\) for \(P^{\mathrm{pre}}\) to \(0.14\,h\,\mathrm{Mpc}^{-1}\), which significantly reduces the bias in \(f\). We find that raising \(k_{\min}\) for \(P^{\mathrm{cross}}\) also yields similar but milder benefits. Ultimately, we opt to remove large-scale modes of \(P^{\mathrm{pre}}\) alone, given its stronger correlation with \(P^{\mathrm{post}}\) and its weaker individual constraining power.

With these adjustments, the joint fit \(P^{\mathrm{all}}\) delivers unbiased best-fit parameter values while improving constraints relative to \(P^{\mathrm{post}}\) alone, yielding uncertainty reductions of about 11\%, 16\%, and 14\% in \(\alpha_{\parallel}\), \(\alpha_{\perp}\), and \(f\), respectively. Additionally, the figure of merit (FoM) for the joint fit is enhanced by 10.5\%. These results corroborate earlier findings \citep{Wang:2022nlx,prepostEmu} that the three spectra together retain complementary information about the pre-reconstruction density field, enabling greater precision than any individual spectrum can achieve. 

We confirm that each power spectrum model separately reproduces the simulation results well, barring some mismatch in the BAO wiggles for \(P^{\mathrm{pre}}\). Likewise, the joint analysis accurately recovers both simulation and theoretical predictions. 

{In this study, we focus on a relatively high redshift slice at \(z = 1.02\). However, it would be valuable to assess our model and methodology at other redshifts in future studies, especially at lower redshifts. At lower redshifts, nonlinear effects are more pronounced, and the improvements from reconstruction tend to be more significant \citep[e.g.,][]{Hikage:2020fte,Wang:2022nlx}. Therefore, we expect that the joint analysis method will be even more effective at lower redshifts, although it is more challenging to apply PT-based models to smaller scales at those redshifts}

Looking ahead to applications in real surveys such as DESI, we plan to improve the theoretical model by recognizing the mismatches between assumed and true cosmological parameters for the BAO reconstruction \citep{Sherwin:2018wbu}, incorporating the IR resummation \citep{Sugiyama:2024eye, Sugiyama:2024qsw}, introducing galaxy bias and other observational effects, and evaluating the approach with halo or galaxy catalogs. 
{We will also explore the FFTLog technique \citep{Hamilton:1999uv} to accelerate the evaluation of loop integrals in SPT, and theory-based emulators \citep{Donald-McCann:2022pac} to further improve the efficiency of the theoretical predictions.}

\begin{acknowledgements}
We thank Chiaki Hikage for contributions in the early stage of this work. WZ, RZ, XM, YW, and GBZ are supported by the National Natural Science Foundation of China (NSFC) under Grant 11925303. KK is supported by the Science and Technology Facilities Council (STFC) under Grant ST/W001225/1. RT is supported by JSPS KAKENHI grant Nos. JP22H00130 and JP20H05855. YW further acknowledges support form the National Key R\&D Program of China No. (2022YFF0503404, 2023YFA1607800, 2023YFA1607803), NSFC Grants (12273048, 12422301), the CAS Project for Young Scientists in Basic Research (No. YSBR-092), and the Youth Innovation Promotion Association CAS. GBZ also acknowledges support from the CAS Project for Young Scientists in Basic Research (No. YSBR-092), the China Manned Space Project, and the New Cornerstone Science Foundation through the XPLORER Prize.
\end{acknowledgements}


\appendix

\section{Supplementary Test Results for Different Scale Ranges}
\label{sec:appendix}

In this section, we provide supplementary results that explore how the choice of scale range affects our analysis. Specifically, we consider a more conservative maximum wavenumber of \(k_{\max}=0.16\,h\,\mathrm{Mpc}^{-1}\), complementing our primary results at \(k_{\max}=0.20\,h\,\mathrm{Mpc}^{-1}\) in the main text. 

\begin{figure}[ht]
\centering
\includegraphics[width=0.8\textwidth]{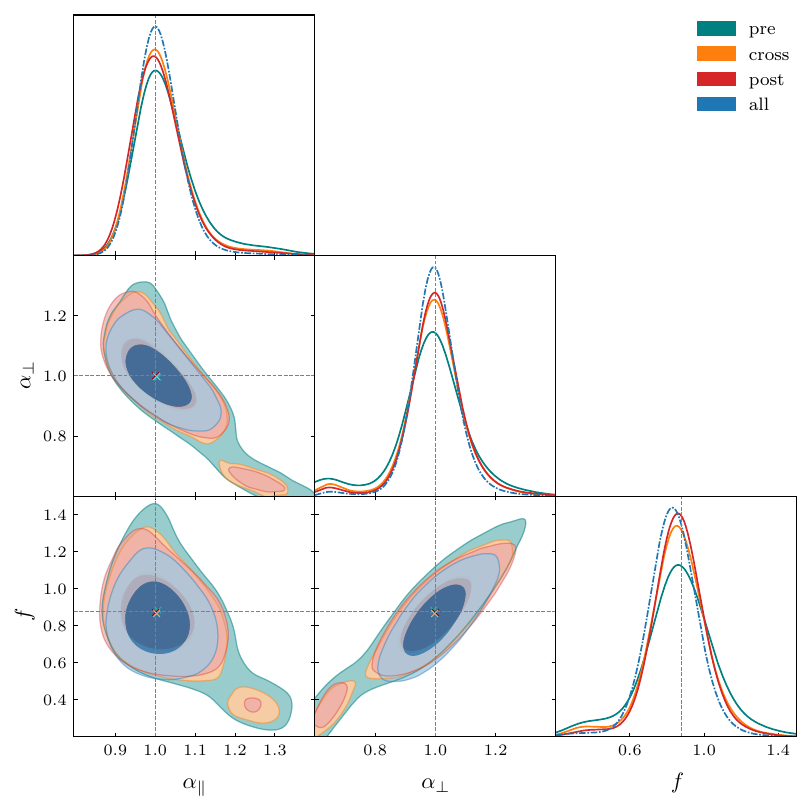}
\caption{\label{fig:contour_0p160} 
Similar to Fig.~\ref{fig:contour_0p200}, but with \(k_{\max}=0.16\,h\,\mathrm{Mpc}^{-1}\).}
\end{figure}

Figure~\ref{fig:contour_0p160} shows the constraints on \(\alpha_{\parallel}\), \(\alpha_{\perp}\), and \(f\) obtained from the individual fits of \(P^{\mathrm{pre}}\), \(P^{\mathrm{post}}\), and \(P^{\mathrm{cross}}\), as well as the joint fit \(P^{\mathrm{all}}\), at \(k_{\max} = 0.16\,h\,\mathrm{Mpc}^{-1}\). Here, \(P^{\mathrm{cross}}\) performs similarly to \(P^{\mathrm{post}}\). The combined fit \(P^{\mathrm{all}}\) continues to yield the tightest constraints—particularly on the AP parameters—but offers only a marginal improvement in \(f\) compared to \(P^{\mathrm{post}}\). This is in contrast to the more substantial gain in \(f\) noted in the main text for \(k_{\max} = 0.20\,h\,\mathrm{Mpc}^{-1}\). The detailed results are presented in Table~\ref{table_0p160}, which also compares the Figure of Merit (FoM) across different scenarios. Specifically, \(P^{\mathrm{all}}\) reduces the uncertainties in \(\alpha_{\parallel}\), \(\alpha_{\perp}\), and \(f\) by 17\%, 17\%, and 8\%, respectively, relative to \(P^{\mathrm{post}}\), while increasing the FoM by 8.3\%.

\begin{table}[ht]
\centering
\caption{Same format as Table~\ref{table_0p200}, but corresponding to Fig.~\ref{fig:contour_0p160} at \(k_{\max}=0.16\,h\,\mathrm{Mpc}^{-1}\).}
\label{table_0p160}
\begin{tabular}{l c c c c c}
\hline\hline
Case & \(P^{\mathrm{pre}}\) & \(P^{\mathrm{post}}\) & \(P^{\mathrm{cross}}\) & \(P^{\mathrm{all}}\) & \(1 - \sigma^\mathrm{all}/\sigma^\mathrm{post}\) \\
\hline

\(\alpha_\parallel\) 
& \(1.029^{+0.045}_{-0.088}\)
& \(1.011^{+0.048}_{-0.073}\)
& \(1.017^{+0.044}_{-0.073}\)
& \(1.010^{+0.045}_{-0.061}\)
& \(17\%\) \\[6pt]

\(\alpha_\perp\)
& \(0.986^{+0.11}_{-0.092}\)
& \(1.00 \pm 0.10\)
& \(0.995^{+0.086}_{-0.077}\)
& \(0.998 \pm 0.083\)
& \(17\%\) \\[6pt]

\(f\)
& \(0.86^{+0.19}_{-0.17}\)
& \(0.86 \pm 0.15\)
& \(0.85 \pm 0.17\)
& \(0.83 \pm 0.14\)
& \(8\%\) \\

\hline
FoM 
& 12.1
& 14.4
& 14.0
& 15.6
& 8.3\% \\

\hline
\end{tabular}
\end{table}

\begin{figure}[htbp]
\centering
\includegraphics[width=0.48\textwidth]{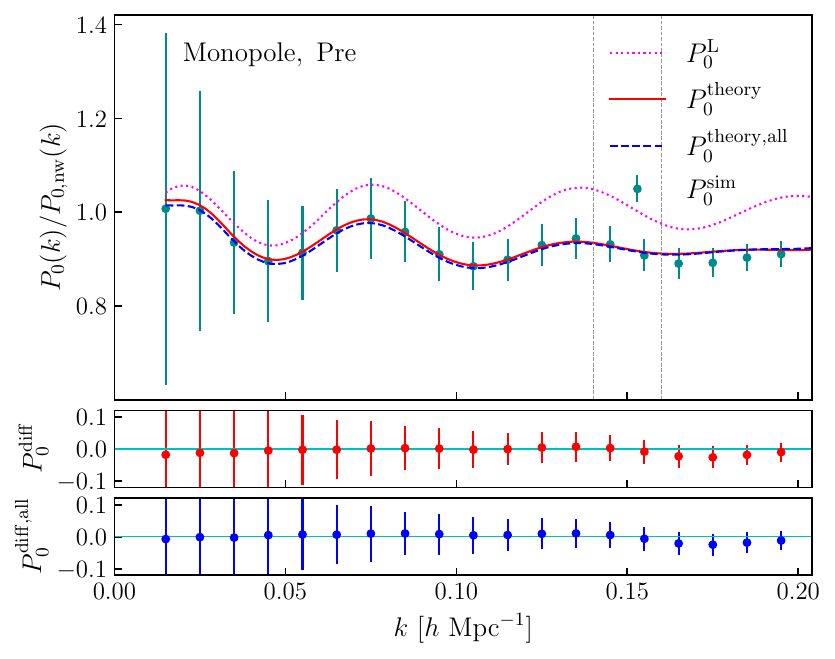}
\includegraphics[width=0.48\textwidth]{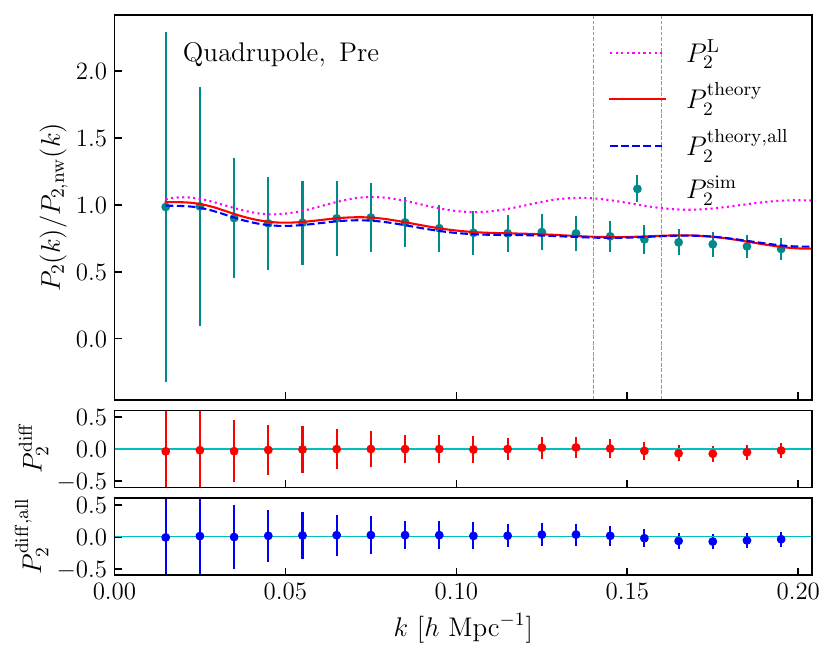}

\includegraphics[width=0.48\textwidth]{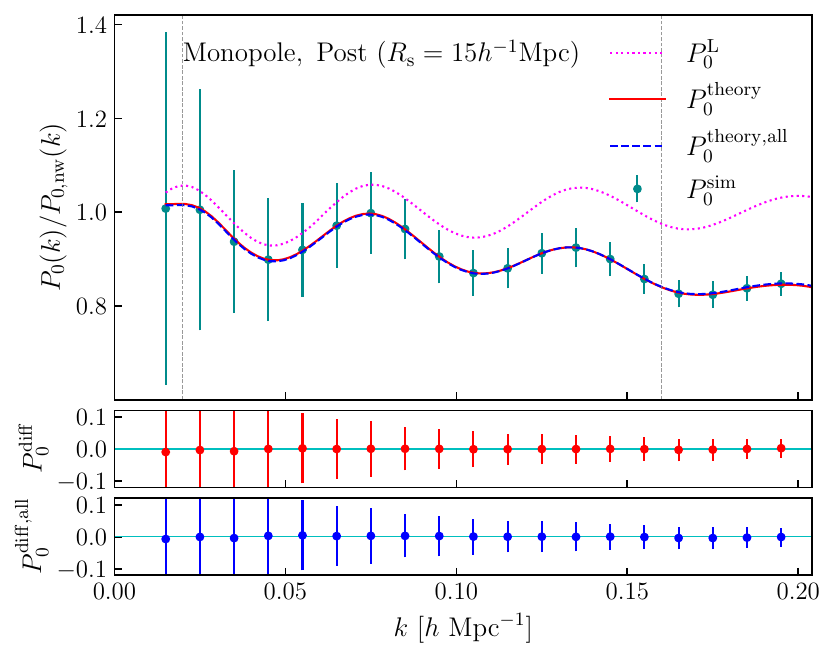}
\includegraphics[width=0.48\textwidth]{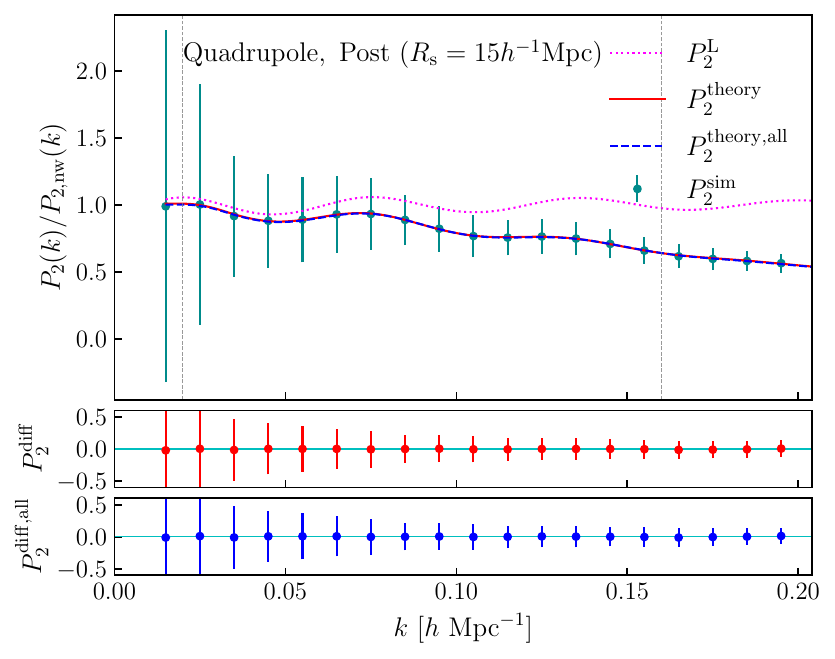}

\includegraphics[width=0.48\textwidth]{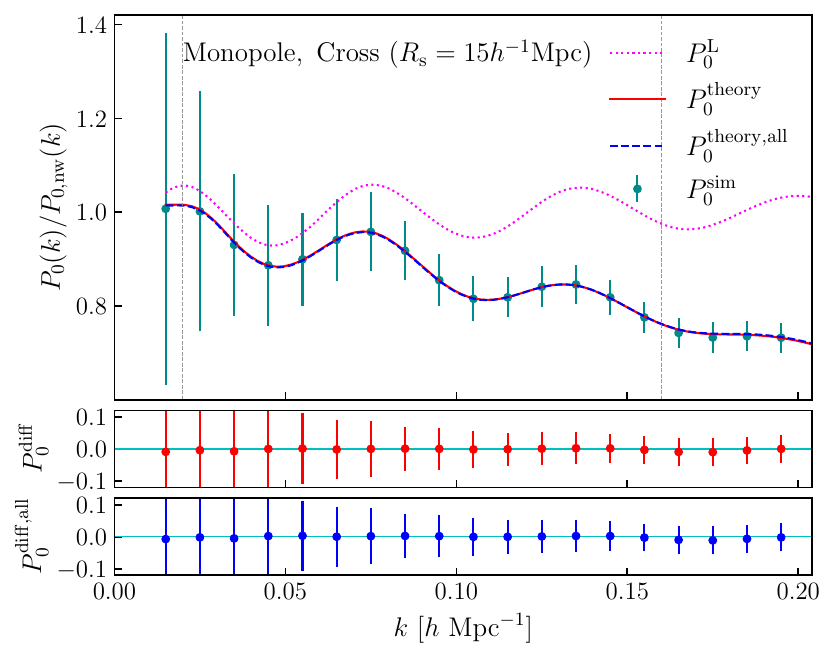}
\includegraphics[width=0.48\textwidth]{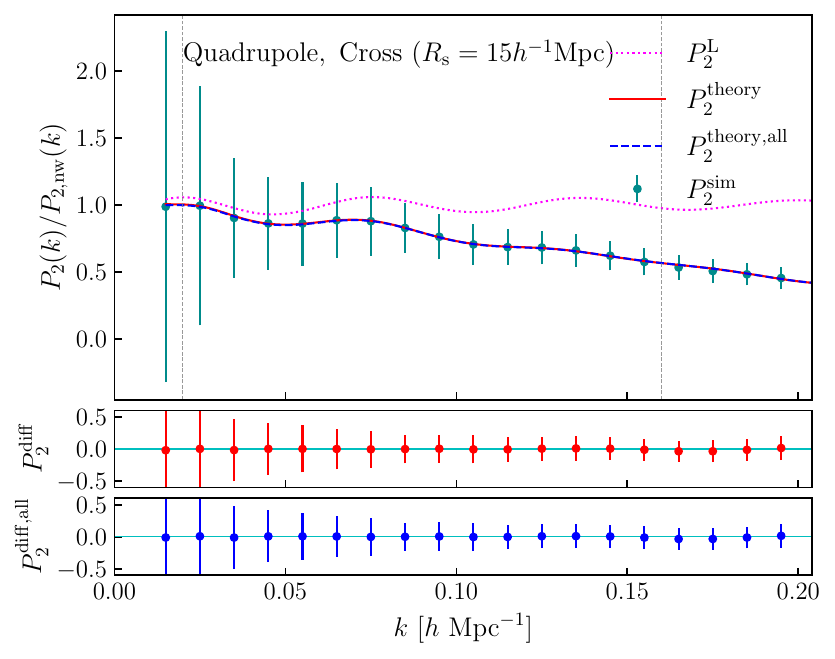}
\caption{\label{fig:wiggle_plk_0p160} 
Same as Fig.~\ref{fig:wiggle_plk_0p200}, but with \(k_{\max}=0.16\,h\,\mathrm{Mpc}^{-1}\). 
Each panel compares the simulation data to theoretical predictions derived from best-fit parameters.}
\end{figure}

In Fig.~\ref{fig:wiggle_plk_0p160}, we compare the theoretical predictions (using best-fit parameters corresponding to Fig.~\ref{fig:contour_0p160}) against the measured multipoles. We again find good agreement between theory and data for both individual fits and the joint fit \(P^{\mathrm{all}}\).

\begin{figure}
\centering
\includegraphics[width=0.49\textwidth]{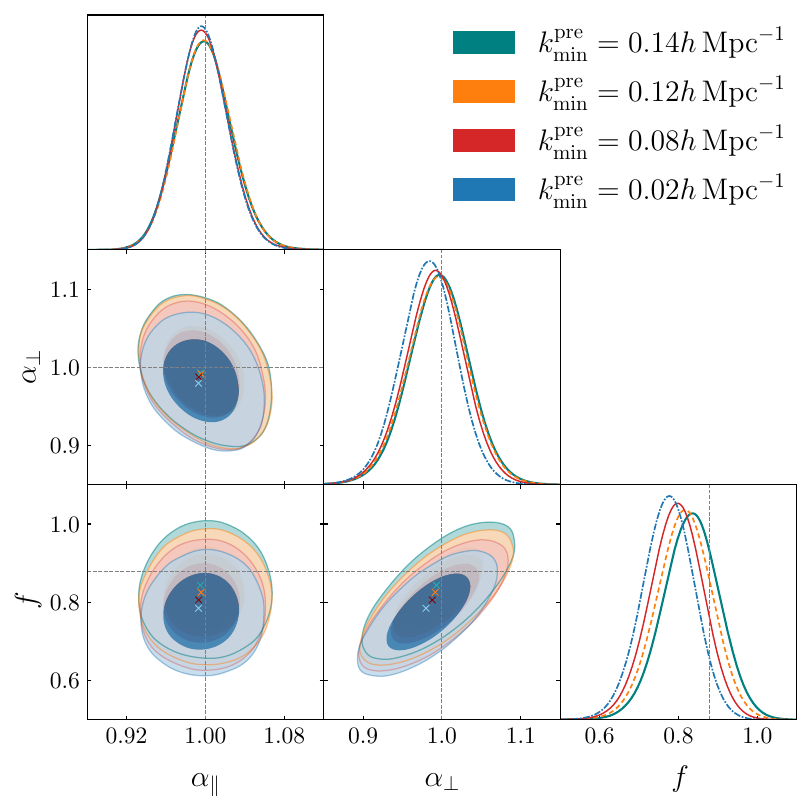}
\includegraphics[width=0.49\textwidth]{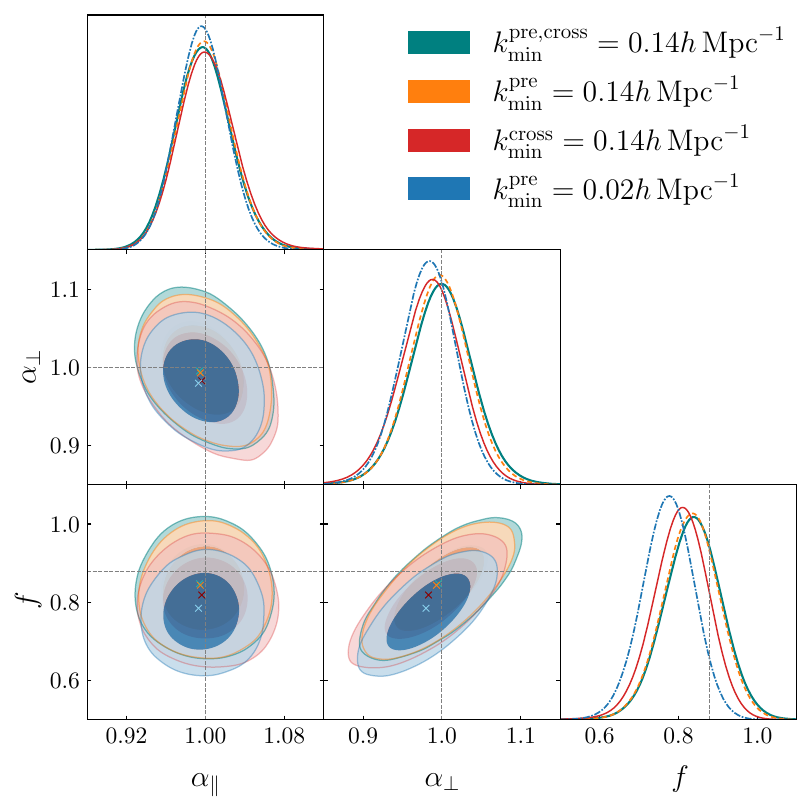}
\caption{\label{fig:kmin_test} 
Constraints on \(\alpha_\parallel\), \(\alpha_\perp\), and \(f\) from joint fits of \(P^{\mathrm{all}}\) under different large-scale cuts, with \(\,R_\mathrm{s} = 15\,h^{-1}\mathrm{Mpc}\) and \(\,k_{\max}^{\mathrm{all}} = 0.20\,h\,\mathrm{Mpc}^{-1}\) held fixed. 
\emph{Left panel}: varying \(k_{\min}^{\mathrm{pre}}=0.08\,h\,\mathrm{Mpc}^{-1}\) (red), \(0.12\,h\,\mathrm{Mpc}^{-1}\) (orange), and \(0.14\,h\,\mathrm{Mpc}^{-1}\) (teal), compared to the baseline \(k_{\min}^{\mathrm{all}}=0.02\,h\,\mathrm{Mpc}^{-1}\) (blue). 
\emph{Right panel}: increasing \(k_{\min}^{\mathrm{cross}}=0.14\,h\,\mathrm{Mpc}^{-1}\) (red), \(k_{\min}^{\mathrm{pre}}=0.14\,h\,\mathrm{Mpc}^{-1}\) (orange), or both simultaneously (teal), again relative to the baseline (blue). 
Crosses mark the best-fit values for each setup, and gray dashed lines denote the fiducial parameters.}
\end{figure}

In Section~\ref{sec:validation} of the main text, we stressed the importance of cutting large-scale modes to ensure robust joint fits of \(P^{\mathrm{all}}\). Here, we expand on that discussion by fixing \(k_{\max}^{\mathrm{all}}=0.20\,h\,\mathrm{Mpc}^{-1}\) and exploring how varying \(k_{\min}\) influences the resulting parameter constraints. As shown in Fig.~\ref{fig:kmin_test}, taking \(k_{\min}=0.02\,h\,\mathrm{Mpc}^{-1}\) for all spectra produces a best-fit linear growth rate \(f\) that is noticeably smaller than its fiducial value, reflecting the strong correlations among the three power spectra on large scales and the ensuing instability in the inverse covariance matrix. 
The left panel of Fig.~\ref{fig:kmin_test} demonstrates that this bias in \(f\) decreases as \(k_{\min}^{\mathrm{pre}}\) is increased. Similarly, the right panel shows that raising \(k_{\min}^{\mathrm{cross}}\), or both \(k_{\min}^{\mathrm{pre}}\) and \(k_{\min}^{\mathrm{cross}}\), mitigates the bias, although to a slightly lesser degree. Based on these tests, in the main text we chose \(k_{\min}^{\mathrm{pre}}=0.14\,h\,\mathrm{Mpc}^{-1}\) for the joint fits, as this approach most effectively reduces biases while retaining sufficient constraining power.

\newpage
\bibliographystyle{raa}

\bibliography{bibtex}

\begin{thebibliography}{67}
\providecommand\natexlab[1]{#1}
\providecommand\JournalTitle[1]{#1}

\bibitem[Adame {et~al.}(2024)]{DESI:2024hhd}
Adame, A.~G., {et~al.} 2024, arXiv:2411.12022

\bibitem[Ade {et~al.}(2016)]{Planck:2015fie}
Ade, P. A.~R., {et~al.} 2016, Astron. Astrophys., 594, A13

\bibitem[Aghamousa {et~al.}(2016)]{DESI:2016fyo}
Aghamousa, A., {et~al.} 2016, arXiv:1611.00036

\bibitem[Alcock \& Paczynski(1979)]{Alcock:1979mp}
Alcock, C., \& Paczynski, B. 1979, Nature, 281, 358

\bibitem[Baumann {et~al.}(2012)]{Baumann:2010tm}
Baumann, D., Nicolis, A., Senatore, L., \& Zaldarriaga, M. 2012, JCAP, 07, 051

\bibitem[Bernardeau {et~al.}(2002)]{Bernardeau:2001qr}
Bernardeau, F., Colombi, S., Gaztanaga, E., \& Scoccimarro, R. 2002, Phys.
  Rept., 367, 1

\bibitem[Chen {et~al.}(2019)]{Chen:2019lpf}
Chen, S.-F., Vlah, Z., \& White, M. 2019, JCAP, 09, 017

\bibitem[Chen {et~al.}(2024)]{Chen:2024exy}
Chen, X., Padmanabhan, N., \& Eisenstein, D.~J. 2024, arXiv:2412.00968

\bibitem[Cole(2005)]{Cole2005}
Cole, S. e.~a. 2005, Monthly Notices of the Royal Astronomical Society, 362,
  505

\bibitem[Crocce {et~al.}(2006)]{Crocce:2006ve}
Crocce, M., Pueblas, S., \& Scoccimarro, R. 2006, Mon. Not. Roy. Astron. Soc.,
  373, 369

\bibitem[Crocce \& Scoccimarro(2008)]{Crocce:2007dt}
Crocce, M., \& Scoccimarro, R. 2008, Phys. Rev. D, 77, 023533

\bibitem[D'Amico {et~al.}(2020)]{DAmico:2019fhj}
D'Amico, G., Gleyzes, J., Kokron, N., {et~al.} 2020, JCAP, 05, 005

\bibitem[Dembinski \& et~al.(2020)]{iminuit}
Dembinski, H., \& et~al., P.~O. 2020

\bibitem[Donald-McCann {et~al.}(2022)]{Donald-McCann:2022pac}
Donald-McCann, J., Koyama, K., \& Beutler, F. 2022, Mon. Not. Roy. Astron.
  Soc., 518, 3106

\bibitem[Eisenstein(2005)]{Eisenstein2005}
Eisenstein, D. J. e.~a. 2005, The Astrophysical Journal, 633, 560

\bibitem[Eisenstein \& Hu(1998)]{Eisenstein1998}
Eisenstein, D.~J., \& Hu, W. 1998, The Astrophysical Journal, 496, 605

\bibitem[Eisenstein {et~al.}(1998{\natexlab{a}})]{Eisenstein:1997gf}
Eisenstein, D.~J., Hu, W., Silk, J., \& Szalay, A.~S. 1998{\natexlab{a}},
  Astrophys. J. Lett., 494, L1

\bibitem[Eisenstein {et~al.}(1998{\natexlab{b}})]{Eisenstein:1998tu}
Eisenstein, D.~J., Hu, W., \& Tegmark, M. 1998{\natexlab{b}}, Astrophys. J.
  Lett., 504, L57

\bibitem[Eisenstein {et~al.}(2007{\natexlab{a}})]{Eisenstein:2006nk}
Eisenstein, D.~J., Seo, H.-j., Sirko, E., \& Spergel, D. 2007{\natexlab{a}},
  Astrophys. J., 664, 675

\bibitem[Eisenstein {et~al.}(2007{\natexlab{b}})]{Eisenstein:2006nj}
Eisenstein, D.~J., Seo, H.-j., \& White, M.~J. 2007{\natexlab{b}}, Astrophys.
  J., 664, 660

\bibitem[Fl\"oss \& Meerburg(2024)]{Floss:2023ylq}
Fl\"oss, T., \& Meerburg, P.~D. 2024, JCAP, 02, 031

\bibitem[Fry(1984)]{Fry:1983cj}
Fry, J.~N. 1984, Astrophys. J., 279, 499

\bibitem[Goroff {et~al.}(1986)]{Goroff:1986ep}
Goroff, M.~H., Grinstein, B., Rey, S.~J., \& Wise, M.~B. 1986, Astrophys. J.,
  311, 6

\bibitem[Hamilton(2000)]{Hamilton:1999uv}
Hamilton, A. J.~S. 2000, Mon. Not. Roy. Astron. Soc., 312, 257

\bibitem[Hartlap {et~al.}(2007)]{Hartlap:2006kj}
Hartlap, J., Simon, P., \& Schneider, P. 2007, Astron. Astrophys., 464, 399

\bibitem[Heavens {et~al.}(1998)]{Heavens:1998es}
Heavens, A.~F., Matarrese, S., \& Verde, L. 1998, Mon. Not. Roy. Astron. Soc.,
  301, 797

\bibitem[Hikage {et~al.}(2017)]{hikage2017perturbation}
Hikage, C., Koyama, K., \& Heavens, A. 2017, Physical Review D, 96, 043513

\bibitem[Hikage {et~al.}(2020{\natexlab{a}})]{Hikage:2019ihj}
Hikage, C., Koyama, K., \& Takahashi, R. 2020{\natexlab{a}}, Phys. Rev. D, 101,
  043510

\bibitem[Hikage {et~al.}(2020{\natexlab{b}})]{Hikage:2020fte}
Hikage, C., Takahashi, R., \& Koyama, K. 2020{\natexlab{b}}, Phys. Rev. D, 102,
  083514

\bibitem[Hinton {et~al.}(2017)]{Hinton:2016atz}
Hinton, S.~R., {et~al.} 2017, Mon. Not. Roy. Astron. Soc., 464, 4807

\bibitem[{Ivanov}(2022)]{ivanov2022EFT}
{Ivanov}, M.~M. 2022, arXiv e-prints, arXiv:2212.08488

\bibitem[Ivanov {et~al.}(2020)]{Ivanov:2019pdj}
Ivanov, M.~M., Simonovi\'c, M., \& Zaldarriaga, M. 2020, JCAP, 05, 042

\bibitem[Jain \& Bertschinger(1994)]{Jain:1993jh}
Jain, B., \& Bertschinger, E. 1994, Astrophys. J., 431, 495

\bibitem[James \& Roos(1975)]{James:1975dr}
James, F., \& Roos, M. 1975, Comput. Phys. Commun., 10, 343

\bibitem[Kaiser(1987)]{Kaiser:1987qv}
Kaiser, N. 1987, Mon. Not. Roy. Astron. Soc., 227, 1

\bibitem[Lewis(2019)]{Lewis:2019xzd}
Lewis, A. 2019, arXiv:1910.13970

\bibitem[Lewis \& Bridle(2002)]{Lewis:2002ah}
Lewis, A., \& Bridle, S. 2002, Phys. Rev. D, 66, 103511

\bibitem[Lewis {et~al.}(2000)]{Lewis:1999bs}
Lewis, A., Challinor, A., \& Lasenby, A. 2000, Astrophys. J., 538, 473

\bibitem[Matsubara(2008{\natexlab{a}})]{matsubara2008nonlinear}
Matsubara, T. 2008{\natexlab{a}}, Physical Review D, 78, 083519

\bibitem[Matsubara(2008{\natexlab{b}})]{Matsubara_2008}
Matsubara, T. 2008{\natexlab{b}}, Physical Review D, 77

\bibitem[Meiksin {et~al.}(1999)]{Meiksin:1998ra}
Meiksin, A., White, M.~J., \& Peacock, J.~A. 1999, Mon. Not. Roy. Astron. Soc.,
  304, 851

\bibitem[Nishimichi {et~al.}(2009)]{Nishimichi:2008ry}
Nishimichi, T., {et~al.} 2009, Publ. Astron. Soc. Jap., 61, 321

\bibitem[Noh {et~al.}(2009)]{Noh_2009}
Noh, Y., White, M., \& Padmanabhan, N. 2009, Physical Review D, 80

\bibitem[Padmanabhan {et~al.}(2009)]{Padmanabhan:2008dd}
Padmanabhan, N., White, M., \& Cohn, J.~D. 2009, Phys. Rev. D, 79, 063523

\bibitem[Schmittfull {et~al.}(2015)]{Schmittfull_2015}
Schmittfull, M., Feng, Y., Beutler, F., Sherwin, B., \& Chu, M.~Y. 2015,
  Physical Review D, 92

\bibitem[Scoccimarro {et~al.}(1999)]{Scoccimarro:1999ed}
Scoccimarro, R., Couchman, H. M.~P., \& Frieman, J.~A. 1999, Astrophys. J.,
  517, 531

\bibitem[Scoccimarro \& Frieman(1996)]{Scoccimarro:1996se}
Scoccimarro, R., \& Frieman, J. 1996, Astrophys. J., 473, 620

\bibitem[Senatore \& Zaldarriaga(2014)]{Senatore:2014vja}
Senatore, L., \& Zaldarriaga, M. 2014, arXiv:1409.1225

\bibitem[Seo {et~al.}(2016)]{Seo:2015eyw}
Seo, H.-J., Beutler, F., Ross, A.~J., \& Saito, S. 2016, Mon. Not. Roy. Astron.
  Soc., 460, 2453

\bibitem[Seo {et~al.}(2008)]{Seo:2008yx}
Seo, H.-J., Siegel, E.~R., Eisenstein, D.~J., \& White, M. 2008, Astrophys. J.,
  686, 13

\bibitem[Seo {et~al.}(2010)]{Seo:2009fp}
Seo, H.-J., Eckel, J., Eisenstein, D.~J., {et~al.} 2010, Astrophys. J., 720,
  1650

\bibitem[Sherwin \& White(2019)]{Sherwin:2018wbu}
Sherwin, B.~D., \& White, M. 2019, JCAP, 02, 027

\bibitem[Shirasaki {et~al.}(2021)]{Shirasaki:2020vkk}
Shirasaki, M., Sugiyama, N.~S., Takahashi, R., \& Kitaura, F.-S. 2021, Phys.
  Rev. D, 103, 023506

\bibitem[Smith {et~al.}(2008)]{Smith:2007gi}
Smith, R.~E., Scoccimarro, R., \& Sheth, R.~K. 2008, Phys. Rev. D, 77, 043525

\bibitem[Springel(2005)]{Springel:2005mi}
Springel, V. 2005, Mon. Not. Roy. Astron. Soc., 364, 1105

\bibitem[Sugiyama(2024{\natexlab{a}})]{Sugiyama:2024eye}
Sugiyama, N. 2024{\natexlab{a}}, Phys. Rev. D, 110, 063528

\bibitem[Sugiyama(2024{\natexlab{b}})]{Sugiyama:2024qsw}
Sugiyama, N. 2024{\natexlab{b}}, arXiv:2403.18262

\bibitem[Sugiyama(2024{\natexlab{c}})]{Sugiyama:2024ggt}
Sugiyama, N. 2024{\natexlab{c}}, arXiv:2406.01001

\bibitem[Torrado \& Lewis(2021)]{Torrado:2020dgo}
Torrado, J., \& Lewis, A. 2021, JCAP, 05, 057

\bibitem[Wang {et~al.}(2024{\natexlab{a}})]{prepostEmu}
Wang, Y., {et~al.} 2024{\natexlab{a}}, Astrophys. J., 966, 35

\bibitem[Wang {et~al.}(2024{\natexlab{b}})]{Wang:2022nlx}
Wang, Y., {et~al.} 2024{\natexlab{b}}, Commun. Phys., 7, 130

\bibitem[White(2015)]{White:2015eaa}
White, M. 2015, Mon. Not. Roy. Astron. Soc., 450, 3822

\bibitem[Zang \& Zhu(2024)]{Zang:2023rpx}
Zang, S.-H., \& Zhu, H.-M. 2024, Astrophys. J., 961, 160

\bibitem[Zel'Dovich(1970)]{zel1970gravitational}
Zel'Dovich, Y.~B. 1970, Astronomy and astrophysics, 5, 84

\bibitem[Zhang {et~al.}(2025)]{pkcross:2025}
Zhang, W., Zhao, R., Mu, X., {et~al.} 2025, arXiv:2502.08186

\bibitem[Zhao {et~al.}(2024)]{Zhao:2023ebp}
Zhao, R., {et~al.} 2024, Mon. Not. Roy. Astron. Soc., 532, 783

\bibitem[Zhu {et~al.}(2018)]{Zhu:2017vtj}
Zhu, H.-M., Yu, Y., \& Pen, U.-L. 2018, Phys. Rev. D, 97, 043502

\end{thebibliography}

\end{document}